\documentclass[11pt,a4paper]{JHEP3_mod2}

\usepackage{amsmath,latexsym,amssymb,slashed}
\usepackage{graphicx}				% inclusion figures
\usepackage{amssymb}					% collection symboles math�matiques
\usepackage{tabularx}				% gestion avanc�e tableaux
\usepackage[vcentermath]{youngtab}	% young tableaux

\newcommand{\eq}[1]{\begin{equation}#1\end{equation}}
\newcommand{\spl}[1]{\begin{split}#1\end{split}}

\newcommand{\slsh}[1]{\displaystyle{\not} #1}

\newcommand{\la}{\lambda}

\newcommand\ga{\gamma}

\newcommand{\boxedeq}[1]{
\begin{equation}
\fbox{
\rule[0.7cm]{0pt}{0pt}
$#1$
\rule[-0.45cm]{0pt}{0pt}
}
\end{equation}
}

\def\d{\text{d}}

\author{Bertrand Sou\`{e}res and Dimitrios Tsimpis\\
Universit\'{e} Claude Bernard (Lyon 1)\\
UMR 5822, CNRS/IN2P3, Institut de Physique Nucl\'{e}aire de Lyon\\
4 rue Enrico Fermi,
F-69622 Villeurbanne Cedex,  France\\

E-mail:
\email{soueres@ipnl.in2p3.fr}, \email{tsimpis@ipnl.in2p3.fr}}
\abstract{
We use  the superspace formulation of (massive) IIA supergravity to obtain the explicit form of the dilatino terms, and  
we find that the quartic-dilatino term is positive.~The theory admits a  ten-dimensional  de Sitter solution, obtained by assuming a nonvanishing  quartic-dilatino condensate which 
generates a positive cosmological constant. Moreover,  in the presence of dilatino condensates,  the theory admits formal  four-dimensional de Sitter solutions of the form $dS_4\times M_6$, 
where  $M_6$ is a six-dimensional K\"{a}hler-Einstein manifold of positive scalar curvature. 
}
\title{De Sitter space from\\ dilatino condensates in (massive) IIA}
\preprint{}

\newcommand{\swed}{{\scriptscriptstyle \wedge}}

\begin{document}
\setlength{\parindent}{0pt}
%\newpage

%%%%%%%%%%%%%%%%%%%%%%%%%%%%%%%%%%%%%%%%%%%%%%%%%%%%%%%%%%%%%%%%%%%%%%%%%%%%%%%%%%%

%%%%%%%%%%%%%%%%%%%%%%%%%%%%%%%%%%%%%%%%DRAFT 01%%%%%%%%%%%%%%%%%%%%%%%%%%%%%%%%%%%%%%%

\section{Introduction and summary}

Fermionic condensates have been considered in the past mostly in the context of  heterotic theory \cite{Dine:1985rz, Derendinger:1985kk, LopesCardoso:2003sp, Derendinger:2005ed, Manousselis:2005xa, Chatzistavrakidis:2012qb, Gemmer:2013ica, Quigley:2015jia, Minasian:2017eur} and, to a lesser extent, in eleven-dimensional supergravity \cite{Duff:1982yi,Jasinschi:1986ze}. Of course spinor vevs must vanish in a 
Lorentz-invariant vacuum, however scalar quadratic- and quartic-fermion condensates are allowed by the symmetry of the vacuum and may be generated by 
quantum effects.

In (massive) type IIA theory there is a single scalar that can be constructed in ten dimensions out of four dilatini, as can be seen by e.g.~the Fierz identities (\ref{fids}) below.  
The presence of a unique quartic-dilatino term in the action thus gives a simple and interesting possibility to generate 
a positive cosmological constant via fermionic condensation.  
As we will see this possibility is indeed realized, in that the quartic-dilatino term of the theory turns out to be positive. Moreover, assuming nonvanishing dilatino condensates, one can obtain both a maximally-symmetric ten-dimensional de Sitter vacuum $dS_{10}$ and a compactification to four dimensional  de Sitter space $dS_4$, of the form $dS_{4}\times M_6$ with $M_6$ a K\"{a}hler-Einstein six-dimensional manifold of positive curvature (such as e.g.~$\mathbb{CP}^3$ with the Fubini-Study metric).

Let us be clear that these are {\it  formal} solutions of (massive) IIA supergravity, obtained by simply assuming nonvanishing values of  the dilatino condensates of the theory. Our apporach is similar to e.g.~\cite{Duff:1982yi}, in that we do not offer
 any concrete scenario or mechanism   for the generation of the dilatino condensate.

The quartic-fermion terms in the massive IIA theory were not computed in \cite{Romans:1985tz}. 
On the other hand  all (massive) IIA supergravities admit a unified superspace formulation, given in \cite{Tsimpis:2005vu}, in which the quartic-fermion terms are given implicitly. 
Unfortunately their explicit form was not worked out in \cite{Tsimpis:2005vu}. 
However it was conjectured in 
\cite{Romans:1985tz} that the quartic-fermion terms are identical in the massive and massless IIA theories. Indeed this follows immediately from the results of \cite{Tsimpis:2005vu}, since at the level of the superspace Bianchi identities given in that reference the massless limit is smooth and 
the quartic-fermion terms do not depend on the mass. 

The massless IIA supergravity theory was first obtained in \cite{Giani:1984wc,Campbell:1984zc,Huq:1983im}  (complete with quartic fermions) by the dimensional reduction of eleven-dimensional supergravity  \cite{Cremmer:1978}. 
Moreover, the quartic-fermion terms of (massive) IIA were given explicitly in \cite{Nicoletti:2011zz}, in the rheonomic formulation. 
These references  could therefore be used in principle to provide the ``missing'' quartic-fermion terms  of Romans supergravity. 
Unfortunately, however, we have been unable to conclude whether 
the quartic-fermion terms in \cite{Giani:1984wc,Campbell:1984zc,Huq:1983im,Nicoletti:2011zz} agree with each other.\footnote{The second author is grateful 
to Stefan Theisen for collaboration on this problem during July-September 2013.}  
Instead in the present paper we derive the dilatino terms from scratch using the superspace formalism of \cite{Tsimpis:2005vu}, and we find agreement with \cite{Giani:1984wc}. 
In deriving these computationally intensive results, we have made extensive use of the computer program \cite{Bertrand} which builds on \cite{Gran:2001:2} to supplement it with various functionalities, 
including explicit spinor indices and their manipulation.

Our strategy will be to first determine the fermionic action, $S_f$, up to gravitino terms. I.e.~$S_f$ is obtained from the full fermionic action by setting the gravitino to zero. 
We will refer to the action thus obtained as the {\it dilatino-condensate action}. The result is given in eq.~(\ref{action2}) below.  Of course setting the 
gravitino to zero is in general inconsistent, since the gravitino couples linearly to terms of the form (flux)$\times$(dilatino) and (dilatino)$^3$. However, in a Lorentz-invariant vacuum, where linear and cubic fermion vevs vanish, this does not lead to an inconsistency.

The dilatino-condensate actions of the present paper should thus be regarded as {\it pseudoactions}: book-keeping devices 
whose variation with respect to the bosonic fields gives the correct bosonic equations of motion in the presence of dilatino condensates (hence their name). Moreover 
the fermionic equations of motion are trivially satisfied in the Lorentz-invariant vacuum.

On the other hand, as we explain in more detail in section \ref{sec:gen}, setting the  gravitino to zero is a frame-dependent statement. Moreover the superspace 
formalism of \cite{Tsimpis:2005vu} turns out to be in a frame different from the Einstein frame. 
Thus the dilatino-condensate action (\ref{action2}) cannot  be compared to the actions in \cite{Giani:1984wc,Campbell:1984zc,Huq:1983im}, which are expressed in the 
Einstein frame, by simply setting the gravitino in those references  to zero.

As we will see in section \ref{sec:gen}, setting the superspace-frame gravitino to zero turns out to be 
equivalent to setting the Einstein-frame gravitino to be proportional to the dilatino. 
Specifically the dilatino-condensate action (\ref{action2}) should be compared with what one obtains from  \cite{Giani:1984wc,Campbell:1984zc,Huq:1983im}  
by imposing (\ref{fv}). This exercise is performed in section \ref{sec:gen}, and we find agreement with \cite{Giani:1984wc}. 
Of course the massive terms in (\ref{action2}) are absent from the massless IIA action of \cite{Giani:1984wc}. Rather they can be compared to what one obtains 
from the  Romans action \cite{Romans:1985tz} by imposing (\ref{fv}), and again we find perfect agreement.

  The generic dilatino-condensate action, obtained by setting to zero the gravitino of an arbitrary frame (parameterized by a real parameter $\beta$), is given in (\ref{action3}) below: it is obtained from the action of \cite{Romans:1985tz} completed with the quartic-fermion terms of \cite{Giani:1984wc}, by imposing (\ref{fvgen}) with arbitrary parameter $\beta$. 
As special cases, the dilatino-condensate actions obtained by setting the Einstein-frame, string-frame gravitino to zero are given in (\ref{action3e}), (\ref{action3s}) respectively.

Having obtained the general dilatino-condensate action, we can look for de Sitter solutions supported by nonvanishing dilatino condensates. 
In section \ref{sec:desitter10} we show that, setting the Einstein-frame gravitino to zero, the theory admits ten-dimensional de Sitter vacua 
supported by the quartic-dilatino condensate, with constant dilaton and vanishing flux.  
In sections  \ref{sec:desitter4sf}, \ref{sec:desitter4} we consider compactifications on six-dimensional K\"{a}hler-Einstein manifolds $M_6$. 
We show that 
setting  the Einstein-frame gravitino to zero leads to four-dimensional de Sitter solutions of the form $dS_4\times M_6$. 
Section \ref{sec:desitter4sf} considers the case of vanishing flux and nonvanishing quadratic- and quartic-dilatino condensates, while 
section \ref{sec:desitter4} considers nonvanishing RR flux and vanishing quadratic-dilatino condensates.

The plan of the remainder of the paper is as follows. The action of (massive) IIA supergravity is obtained in section \ref{sec:massive}, up to gravitino terms. The 
derivation of the bosonic terms, already worked out in \cite{Tsimpis:2005vu}, is reviewed in section \ref{sec:bt}.  In particular we recover the fact that the superspace formalism of \cite{Tsimpis:2005vu} is naturally formulated in a frame different from the Einstein frame. The dilatino terms are derived in section 
\ref{sec:ft}. 
In section \ref{sec:gen} we compare the dilatino-condensate action (\ref{action2}) with what one would obtain from  \cite{Romans:1985tz} and the quartic term in \cite{Giani:1984wc} by 
setting the superspace-frame gravitino to zero, and we find perfect agreement. Furthermore we derive the generic dilatino-condensate action (\ref{action3}) obtained by setting to zero 
the gravitino of an arbitrary frame. In section \ref{sec:desitter} we show that the dilatino condensate action (\ref{action3e}), obtained by setting the Einstein-frame gravitino to zero, admits 
de Sitter vacua of the from $dS_{10}$ and $dS_4\times M_6$, supported by the quartic-dilatino condensate. We conclude in section \ref{sec:conclusions}. Appendix \ref{sec:sse} works out the supersymmetry transformations, while appendix \ref{sec:comparison} compares some different conventions in the literature.

\section{Massive IIA in superspace}\label{sec:massive}

Massless IIA supergravity \cite{Giani:1984wc,Campbell:1984zc,Huq:1983im} was first obtained by the dimensional reduction of eleven-dimensional supergravity  \cite{Cremmer:1978}. 
The massive deformation of IIA supergravity, which cannot be obtained by reduction from eleven dimensions,  
was introduced by Romans in \cite{Romans:1985tz}. Moreover all (massive) IIA supergravities admit a unified superspace formulation, given in \cite{Tsimpis:2005vu}.\footnote{The solution of the superspace Bianchi identities up to mass dimension-1 was previously given in \cite{Carr:1986tk}.}

Here we are interested 
 in giving a nonzero expectation value to the dilatino condensate in (massive) IIA supergravity.   For that purpose we need 
to know the terms both quadratic and quartic in the dilatino. However the quartic-fermion terms (although implicit in  \cite{Tsimpis:2005vu}) were not explicitly derived in that reference.

In this section we will  use the superspace formulation of the theory to extract the explicit form of the 
 dilatino terms in the action. More precisely, Romans supergravity is obtained by setting 
\eq{
L=\frac{3}{4} (\mu\la)+\frac12 me^{2\phi} ~;~~~
L'=-L
~,}
in all equations of \cite{Tsimpis:2005vu}.

\subsection{The bosonic terms}\label{sec:bt}

The bosonic equations of motion appear at mass dimension-2 and are given in eq.~(3.91) of \cite{Tsimpis:2005vu}.  Setting all fermionic superfields to zero and restricting to the $x$-space component of the bosonic superfields (i.e. the lowest-order term in the theta-expansion), the content of the $\mathcal{B}=\mathcal{C}=0$ equations of \cite{Tsimpis:2005vu} can be seen to be equivalent to the following set of equations, 
\eq{\spl{\label{b1}
\d  L_{(2)} +\frac{18i}{5}m e^{2\phi} K_{(3)}&=0\\
i\d  L_{(4)} -\frac{2}{3} K_{(3)}\swed  L_{(2)}
+4 K_{(1)}\swed L_{(4)}&=0\\
i\d \star L_{(4)} +8 K_{(1)}\swed \star L_{(4)}
+24 K_{(3)}\swed  L_{(4)}&=0\\
i\d \star L_{(2)} +12 K_{(1)}\swed \star L_{(2)}
+864K_{(3)}\swed \star L_{(4)}&=0
~.}}
The $\mathcal{A}=\mathcal{D}=0$ equations of \cite{Tsimpis:2005vu} can be seen to be equivalent to, 
\eq{\spl{\label{b2}
\d  K_{(1)} &=0\\
i\d  K_{(3)} -4 K_{(1)}\swed  K_{(3)}
&=0\\
i\d \star K_{(3)} -8 K_{(1)}\swed \star K_{(3)}
-\frac{128}{3} L_{(2)}\swed \star L_{(4)}-768 L_{(4)}\swed  L_{(4)} -\frac{8}{45}me^{2\phi}\star L_{(2)}&=0\\
i\d \star K_{(1)} -12 K_{(1)}\swed \star K_{(1)}
-\frac{32}{3} L_{(2)}\swed \star L_{(2)}- 144K_{(3)}\swed \star K_{(3)}  
-4608 L_{(4)}\swed  \star L_{(4)}+\frac{2}{5}m^2e^{4\phi}&=0
~,}}
together with the Einstein equation,
\eq{\spl{\label{e1}
R_{mn}&=g_{mn}\Big(\frac{3i}{2} \nabla\cdot K_{(1)}+18K_{(1)}^2-\frac{1}{25}m^2e^{4\phi}\Big)\\
&+12i\nabla_{(m}K_{n)}-16K_mK_n
-\frac{64}{9}\Big(  2L^2_{(2)mn} -\frac{1}{8} g_{mn}  L_{(2)}^2 \Big)
\\
&+48\Big(  3K^2_{(3)mn} -\frac{1}{4} g_{mn}  K_{(3)}^2 \Big)
-768\Big(   4L^2_{(4)mn} -\frac{3}{8} g_{mn}  L_{(4)}^2 \Big)
~,}}
where we have set $\Phi_{(p)}^2:=\Phi_{m_1\dots m_p}\Phi^{m_1\dots m_p}$, $\Phi^2_{(p)mn}:=\Phi_{mm_2\dots m_p}\Phi_n{}^{m_2\dots m_p}$, for any $p$-form $\Phi$. Moreover in order to put the Einstein equation in the form (\ref{e1}) we have made use of the last equation in (\ref{b2}).
 Note that the latter  can be  obtained by  acting on the equations of motion of the fermionic superfield, cf.~(\ref{f}) below, with
a spinor derivative and contracting the free spinor indices with each other.

The first equation in (\ref{b2}) above can be solved by introducing a scalar field $\phi$,
\eq{\label{sc} K_{(1)}=\frac{i}{2}\d\phi~,}
where the normalization has been chosen so that $\phi$ is identified with the dilaton. 
The equations above  are not automatically expressed in the Einstein frame in ten dimensions. To transform to the Einstein frame 
we define a new Weyl-rescaled metric,
\eq{\label{rsm}\hat{g}_{mn} =
e^{\frac{3}{2}\phi}{g}_{mn} 
~. }
The Einstein equation then takes the form,
\eq{\spl{\label{e2}
\hat{R}_{mn}&=-\frac{1}{2}\partial_m\phi\partial_n\phi-\frac{1}{25}m^2e^{4\phi}g_{mn}
-\frac{64}{9}\Big(  2L^2_{(2)mn} -\frac{1}{8} g_{mn}  L_{(2)}^2 \Big)
\\
&+48\Big(  3K^2_{(3)mn} -\frac{1}{4} g_{mn}  K_{(3)}^2 \Big)
-768\Big(   4L^2_{(4)mn} -\frac{3}{8} g_{mn}  L_{(4)}^2 \Big)
~,}}
where $\hat{R}_{mn}$ is the Ricci tensor of $\hat{g}$; the contractions on the right-hand side  are taken with respect to the 
unrescaled metric $g$.

The equations above can be recognized as the bosonic equations of  Romans supergravity. 
For example one can readily make contact with the  formulation of \cite{Lust:2004ig}\footnote{\label{f2}We are using 
the conventions of \cite{Lust:2004ig} where all (bosonic) forms are given in 
``superspace conventions'',
\eq{
\Phi_{(p)}=\frac{1}{p!}\Phi_{m_1\dots m_p}\d x^{m_p}\swed\dots\swed\d x^{m_1}~;~~~
\d\Big( \Phi_{(p)} \swed\Psi_{(q)}\Big)=\Phi_{(p)} \swed\d\Psi_{(q)}
+(-1)^q\d\Phi_{(p)} \swed\Psi_{(q)}~.\nonumber
}
These are better suited for our discussion here which derives from the superspace formulation of IIA supergravity in which these conventions are 
the natural ones. 
The Hodge star is defined as follows,
\eq{
\star (\d x^{a_1}\wedge\dots\wedge \d x^{a_p})=\frac{1}{(10-p)!}\varepsilon^{a_1\dots a_p}{}_{b_1\dots b_{10-p}} \d x^{b_1}\wedge\dots\wedge \d x^{b_{10-p}}
~,\nonumber}
so that,
\eq{
\Phi_{(p)}\wedge\star\Phi_{(p)}
=(\star1)\frac{1}{p!}\Phi_{m_1\dots m_p} \Phi^{m_1\dots m_p}
~.\nonumber}
} 
by using the following dictionary,
\eq{\spl{\label{dic}
L_{(2)}&=-\frac{3}{16}F~;~~~
K_{(3)}=-\frac{i}{24}e^{-2\phi}H~;~~~
L_{(4)}=\frac{1}{192}e^{-2\phi}G\\
m&=\frac{5}{2}m^{\rm{there}}
~;~~~\hat{g}_{mn}=g^{\rm{there}}_{mn}
~;~~~
\hat{R}=-R^{\rm{there}}
~,}}
up to fermionic bilinear terms which will be determined in the following, cf.~(\ref{hs}) below. 
The equations of motion read,
\eq{\spl{\label{beom1}
0&=\hat{\nabla}^2\phi-\frac{3}{8}e^{3\phi/2}F^2+\frac{1}{12}e^{-\phi}H^2-\frac{1}{96}e^{\phi/2}G^2 -\frac{4}{5}m^2e^{5\phi/2}\\
0&=\d(e^{3\phi/2}\hat{\star} F)+e^{\phi/2}H\swed \hat{\star} G\\
0&=\d(e^{-\phi}\hat{\star} H)+e^{\phi/2}F\swed\hat{\star} G-\frac{1}{2}G\swed G+\frac{4}{5}m e^{3\phi/2}\hat{\star} F
\\
0&=\d(e^{\phi/2}\hat{\star} G)-H\swed G
~,
}}
where the covariant derivative $\hat{\nabla}$ and the Hodge star $\hat{\star}$  are taken with respect to the 
rescaled metric $\hat{g}$,  
and,
\eq{\spl{\label{beom2}
0=\hat{R}_{mn}&+\frac{1}{2}\partial_m\phi\partial_n\phi+\frac{1}{25}m^2e^{5\phi/2}\hat{g}_{mn}
+\frac{1}{4}e^{3\phi/2}\Big(  2F^2_{(2)mn} -\frac{1}{8} \hat{g}_{mn}  F_{(2)}^2 \Big)\\
&+\frac{1}{12}e^{-\phi}\Big(  3H^2_{(3)mn} -\frac{1}{4} \hat{g}_{mn}  H_{(3)}^2 \Big)
+\frac{1}{48}e^{\phi/2}\Big(   4G^2_{(4)mn} -\frac{3}{8} \hat{g}_{mn}  G_{(4)}^2 \Big)
~,
}}
where 
the contractions on the right-hand side  are computed using $\hat{g}$. 
Moreover the forms obey the following Bianchi identities,
\eq{\label{bi}
\d F=\frac{4}{5}m H~;~~~\d H=0~;~~~\d G=H\wedge F
~.}
It can also be checked that the equations of motion integrate to the following bosonic action in the Einstein frame, cf.~(2.1) of  \cite{Lust:2004ig},
\eq{\spl{\label{ba}S_b=\int\d^{10}x\sqrt{\hat{g}}\Big(
\hat{R}+\frac12 (\partial\phi)^2&+\frac{8}{25}m^2e^{5\phi/2}+\frac{1}{2\cdot 2!}e^{3\phi/2}F^2+\frac{1}{2\cdot 3!}e^{-\phi}H^2+\frac{1}{2\cdot 4!}e^{\phi/2}G^2
\Big) 
+\mathrm{CS}
~,
}}
where  contractions are taken with respect to the rescaled  metric $\hat{g}$ and CS denotes the Chern-Simons term.

\subsection{The dilatino terms}\label{sec:ft}

As explained in the introduction, we are interested in determining  the fermionic action up to gravitino terms. 
The fermionic equations of motion appear at dimension-3/2 and are given in eq.~(4.25) of \cite{Tsimpis:2005vu},
\eq{\spl{\label{f}
i\slsh \nabla \lambda&=-\frac{24}{5}me^{2\phi}\mu-\frac{36}{5}(\mu\la) \mu-\frac{16}{3}L_{(2)}(\gamma^{(2)}\mu)\\
&-12K_{(1)}(\ga^{(1)}\la)
+3K_{(3)}(\ga^{(3)}\la)
+\frac{3}{40}(\mu\ga_{(3)}\mu)(\ga^{(3)}\la)\\
i\slsh \nabla \mu&=\frac{24}{5}me^{2\phi}\lambda+\frac{36}{5}(\mu\la) \la-\frac{16}{3}L_{(2)}(\ga^{(2)}\la)\\
&-12K_{(1)}(\ga^{(1)}\mu)
-3K_{(3)}(\ga^{(3)}\mu)
+\frac{3}{40}(\la\ga_{(3)}\la)(\ga^{(3)}\mu)~.
}}
These are exact superfield equations, i.e. valid to all orders in the theta-expansion. 

In order to identify the fermionic part of the action giving rise to these equations we must first address the following two issues: Firstly, once the fermionic superfields are turned on, the bosonic equations (\ref{b1}),  (\ref{b2}),  (\ref{e1}) will be violated by terms quadratic and quartic in the fermion superfields. In other words, these equations are not valid as full-fledged superspace equations for superfields. In particular, the superspace Bianchi identities for the superforms at mass dimension-1, read:
\eq{\spl{\label{214}
0 \;&=\; \d \hat{K}_1 \\
0 \;&=\; \d \hat{L}_2 + \frac{18}{5} \; m e^{2 \phi} \; \hat{K}_3 \\
0 \;&=\; \d \hat{K}_3 + 4i \; \hat{K}_1 \swed \hat{K}_3 \\
0 \;&=\; \d \hat{L}_4 + \frac{2i}{3} \; \hat{L}_2 \swed \hat{K}_3 - 4i \; \hat{K}_1 \swed \hat{L}_4~,
}}
where the hatted superfields differ in general from the unhatted ones  by spinor superfield bilinears. 
Explicitly in components the Bianchi identities read:
\eq{\spl{
0 \;&=\; D_{[A} \hat{K}_{B)} +\frac{1}{2} {T_{AB}}^{F} \hat{K}_{F}  \\
0 \;&=\; D_{[A} \hat{L}_{BC)} + {T_{[AB|}}^{F} \hat{L}_{F|C)} + \frac{6}{5} \; m e^{2 \phi} \; \hat{K}_{ABC} \\
0 \;&=\; D_{[A} \hat{K}_{BCD)} +\frac{3}{2} {T_{[AB|}}^{F} \hat{K}_{F|CD)} + 4i \; \hat{K}_{[A} \hat{K}_{BCD)} \\
0 \;&=\; D_{[A} \hat{L}_{BCDE)} +2 {T_{[AB|}}^{F} \hat{L}_{F|CDE)} + \frac{4i}{3} \; \hat{L}_{[AB} \hat{K}_{CDE)} - 4i \; \hat{K}_{[A} \hat{L}_{BCDE)}~.
}}
These can be solved following the standard procedure,  taking into account the expressions for the torsion superfield components of \cite{Tsimpis:2005vu}.  
The solution reads,
\eq{\spl{\label{hs}
\hat{K}_{a} &= K_{a}   \\
\hat{L}_{a b} &= L_{a b} + \frac{3}{8} \; \mu \gamma_{ab} \lambda    \\
\hat{K}_{a b c} &= K_{abc}  - \frac{1}{8} \; \mu \gamma_{abc} \mu + \frac{1}{8} \; \lambda \gamma_{abc} \lambda  \\
\hat{L}_{a b c d} &= L_{abcd} + \frac{1}{32} \; \mu \gamma_{abcd} \lambda \; ,
}}
for the top (bosonic) components and

\vskip -.7cm

\begin{minipage}[t]{0.1\linewidth}
\begin{align*}
\hat{K}_{\alpha} &= \frac{i}{2} \; \lambda_{\alpha} \\
\hat{K}^{\alpha} &= \frac{i}{2} \; \mu^{\alpha} \\
\end{align*}
\end{minipage}
\hfill
\begin{minipage}[t]{0.1\linewidth}
\begin{align*}
\hat{L}_{\alpha}{}^{\beta} &= -\frac{3}{16} \; \delta_{\alpha}^{\beta} \\
\hat{L}^{\alpha}{}_{\beta} &= -\frac{3}{16} \; \delta^{\alpha}_{\beta}
\end{align*}
\end{minipage}
\hfill
\begin{minipage}[t]{0.1\linewidth}
\begin{align*}
\hat{K}_{a b \alpha} &= \frac{i}{12} \; (\gamma_{ab} \lambda)_{\alpha} \\
\hat{K}_{a b}{}^{\alpha} &= -\frac{i}{12} \; (\gamma_{ab} \mu)^{\alpha} \\
\hat{K}_{a \alpha \beta} &= -\frac{1}{24} \; (\gamma_{a})_{\alpha \beta} \\
\hat{K}_{a}{}^{\alpha \beta} &= \frac{1}{24} \;(\gamma_{a})^{\alpha \beta}
\end{align*}
\end{minipage}
\hfill
\begin{minipage}[t]{0.1\linewidth}
\begin{align*}
\hat{L}_{a b c \alpha} &= \frac{i}{96} \; (\gamma_{abc} \mu)_{\alpha} \\
\hat{L}_{abc}{}^{\alpha} &= -\frac{i}{96} \; (\gamma_{abc} \lambda)^{\alpha} \\
\hat{L}_{a b \alpha}{}^{\beta} &= -\frac{1}{192} \; {(\gamma_{ab})_{\alpha}}^{\beta} \\
\hat{L}_{ab}{}^{\alpha}{}_{\beta} &= \frac{1}{192} \; {(\gamma_{ab})^{\alpha}}_{\beta}~,
\end{align*}
\end{minipage}

for the remaining components. {\it The ordinary  bosonic  forms are identified with the lowest-order components in the theta-expansion 
of the hatted superfields in} (\ref{hs}).

Secondly, note that the following combinations,
\eq{\spl{\label{mon}
\Delta{T}^{\alpha}&:=-\tilde{T}^{\alpha}+
\frac{344}{225}L\mu
+\frac{8}{9}L_{(2)}(\gamma^{(2)}\mu)
+\frac{8}{45}L_{(4)}(\gamma^{(4)}\mu)\\
&+\frac{8}{9}K_{(1)}(\gamma^{(1)}\lambda)
-\frac{16}{45}K_{(3)}(\gamma^{(3)}\lambda)
-\frac{11}{450}(\mu\gamma_{(3)}\mu)(\ga^{(3)}\lambda)
\\
\Delta{T}_{\alpha}&:=-\tilde{T}_{\alpha}
-\frac{344}{225}L\lambda
+\frac{8}{9}L_{(2)}(\gamma^{(2)}\lambda)
-\frac{8}{45}L_{(4)}(\gamma^{(4)}\lambda)\\
&+\frac{8}{9}K_{(1)}(\gamma^{(1)}\mu)
+\frac{16}{45}K_{(3)}(\gamma^{(3)}\mu)
-\frac{11}{450}(\lambda\ga_{(3)}\la)(\gamma^{(3)}\mu)
~,}}
vanish on-shell, cf.~(4.9),(4.10) of \cite{Tsimpis:2005vu}. 
Hence  we are free to add to the right-hand sides of equations (\ref{f}) terms proportional to $\Delta{T}$ above. When integrated to 
a fermionic action, they induce terms proportional to $\tilde{T}^{\alpha}\lambda_{\alpha}$, $\tilde{T}_{\alpha}\mu^{\alpha}$. Given that $\tilde{T}$ 
is the trace of 
the dimension-3/2 
torsion, these are gravitino terms which we set to zero here.\footnote{The  precise relation between 
$T^{\alpha}_{ab}$ and the gravitino can be derived using the procedure described in detail in e.g.~\cite{Tsimpis:2004gq}  and it is of the form: 
$e_m{}^ae_n{}^bT^{\alpha}_{ab}=\nabla_{[m}\psi^{\alpha}_{n]}+\mathcal{O}(\psi)$. In particular it vanishes upon setting $\psi^{\alpha}_{m}\equiv0$.}

Let us take as our starting point  the fermionic equations (\ref{f}), adding to the right-hand sides the terms $c_1\Delta{T}^{\alpha}$, $c_2\Delta{T}_{\alpha}$,  
as explained in the previous paragraph, 
for some coefficients $c_1$, $c_2$. Provided we take $c_2=c_1$, the resulting equations can be integrated into the following fermionic action:
\eq{\spl{S_f
=&
\int\d^{10}x\sqrt{\hat{g}}e^{(6-8c_1/9)\phi}\Big\{
(\bar{\Lambda}\Gamma^m\nabla_m\Lambda)
-\frac{4}{225}(270-43c_1)e^{5\phi/4}m(\bar{\Lambda}\Lambda)\\
&-(1-\frac{1}{6}c_1)e^{3\phi/4} F_{mn}(\bar{\Lambda}\Gamma^{mn}\Gamma_{11}\Lambda)
+(\frac{1}{8}-\frac{2}{135}c_1)e^{-\phi/2} H_{mnp}(\bar{\Lambda}\Gamma^{mnp}\Gamma_{11}\Lambda)\\
&+\frac{1}{1080}c_1 e^{\phi/4} G_{mnpq}(\bar{\Lambda}\Gamma^{mnpq}\Lambda)
+\frac{2}{5}(15-c_1)(\bar{\Lambda}\Lambda)^2
\Big\}
~,}}
where 
the Dirac gamma-matrices $\Gamma^m$ and the Majorana fermions $\Lambda$ are given in (\ref{gf}), (\ref{gfd}) respectively; 
we have expressed the final result in terms of the rescaled metric (\ref{rsm}) and the bosonic forms in (\ref{dic}), 
with the understanding that the  unhatted forms therein are now replaced by the correponding hatted ones given in (\ref{hs}): 
\eq{\label{dicb}
F:=-\frac{16}{3}\hat{L}_{(2)}~;~~~
H:=24ie^{2\phi}\hat{K}_{(3)}~;~~~
G:=192e^{2\phi} \hat{L}_{(4)}
~.}
The total action (up to gravitino terms) is thus given by:  
%
%\eq{\label{action1} 
$S=S_b+\alpha S_f$, 
%~,}
%
for some coefficient $\alpha$ to be determined. 

Next consider the dilaton  equation of motion,
\eq{\spl{\label{dil2}
0&=-2i\nabla\cdot K_{(1)}-24K_{(1)}^2-\frac{4}{5}m^2e^{4\phi}-\frac{32}{3}\hat{L}_{(2)}^2-48\hat{K}_{(3)}^2
-384\hat{L}_{(4)}^2\\
&-\frac{16}{5}me^{2\phi}(\lambda\mu)-8(\lambda\ga^{(2)}\mu)\hat{L}_{(2)}
+8\big[(\lambda\ga^{(3)}\lambda)-(\mu\ga^{(3)}\mu)\big]\hat{K}_{(3)}
-32(\lambda\ga^{(4)}\mu)\hat{L}_{(4)}+144(\lambda\mu)^2
~,}}
which is an exact superfield equation obtained from the Bianchi identities at dimension-2; it reduces to the bosonic dilaton equation 
given in (\ref{beom1}) upon setting to zero the fermionic superfields, and transforming to the Einstein-frame metric. 
As explained above, we can modify equation (\ref{dil2}) by  adding on the right hand-side  
a term of the form $c_3\lambda_{\alpha}\Delta{T}^{\alpha}+c_4\mu^{\alpha}\Delta{T}_{\alpha}$,  
which vanishes on-shell. This will generate gravitino terms $\lambda_{\alpha}\tilde{T}^{\alpha}$, $\mu^{\alpha}\tilde{T}_{\alpha}$, which we can then set to zero. 
Demanding that the resulting equation of motion coincides with the dilaton equation coming from $S_b+\alpha S_f$, gives an 
overdetermined system of equations for the unknown coefficients $\alpha$, $c_1,\dots, c_4$. The solution reads,
\eq{\alpha=-80~;~~~c_1=c_2=\frac{27}{4}~;~~~c_3=c_4=-45~.}
Plugging back the above into the action we obtain,
%
%\boxedeq{\label{action1}S=S_b-80S_f~,}
%
%
\eq{\spl{\label{action2}
S&=S_b-80
\int\d^{10}x\sqrt{\hat{g}} \Big\{
(\bar{\Lambda}\Gamma^m\nabla_m\Lambda)
+\frac{9}{25}e^{5\phi/4}m(\bar{\Lambda}\Lambda)\\
&+\frac{1}{8}e^{3\phi/4} F_{mn}(\bar{\Lambda}\Gamma^{mn}\Gamma_{11}\Lambda)
+\frac{1}{40}e^{-\phi/2} H_{mnp}(\bar{\Lambda}\Gamma^{mnp}\Gamma_{11}\Lambda)\\
&+\frac{1}{160}e^{\phi/4} G_{mnpq}(\bar{\Lambda}\Gamma^{mnpq}\Lambda)
+\frac{33}{10}(\bar{\Lambda}\Lambda)^2
\Big\}
~,}}
where the bosonic part of the action $S_b$ was given in (\ref{ba}).

The Einstein equation can be used as a further  consistency check. The dimension-2 superspace Bianchi identities give,
\eq{\spl{\label{eeq}
R_{bc}&=
\eta_{bc}\Big(
-\frac{1}{25}m^2e^{4\phi}+\frac{3i}{2}\nabla\cdot K_{(1)}+18K_{(1)}^2
+\frac{8}{9}\hat{L}_{(2)}^2-12\hat{K}_{(3)}^2+288\hat{L}_{(4)}^2\\
&~~~~~~~~~~\! -\frac{36}{5}(\la\mu)me^{2\phi}
-\frac{16}{3}(\la\ga^{(2)}\mu)\hat{L}_{(2)}
+6\big[(\la\ga^{(3)}\la)-(\mu\ga^{(3)}\mu)\big]\hat{K}_{(3)}\\
&~~~~~~~~~~\! +24(\la\ga^{(4)}\mu)\hat{L}_{(4)}
-108(\la\mu)^2
\Big)\\
&+{12i}\nabla_{(b}K_{c)}-16 K_bK_c
-\frac{128}{9}\hat{L}_{(2)bc}^2+144\hat{K}_{(3)bc}^2
-3072\hat{L}_{(4)bc}^2
\\
&+4i(\la\ga_{(b}\nabla_{c)}\la)+4i(\mu\ga_{(b}\nabla_{c)}\mu)
-\frac{32}{3}(\la\ga_{(b}{}^i\mu)\hat{L}_{c)i}
-36\big[(\la\ga_{(b}{}^{ij}\la)-(\mu\ga_{(b}{}^{ij}\mu)\big]\hat{K}_{c)ij}\\
&-192(\la\ga_{(b}{}^{ijk}\mu)\hat{L}_{c)ijk}
~.}}
Proceeding as before,  we note that the following terms vanish on-shell,
\eq{\spl{
\Delta T_a^{\alpha}&:=-\widetilde{T}_a^{\alpha}-\frac{3i}{20}(\ga_a^{(1)} \nabla_{(1)} \mu)-\frac{1}{5}L_{(2)}(\ga_{a}^{(2)}\la)
+\frac{2}{5}L_{(4)}(\ga_{a}^{(4)}\la)\\
&+\frac{1}{5}K_{(1)}(\ga_{a}^{(1)}\mu)-\frac{3}{20}K_{(3)}(\ga_{a}^{(3)}\mu)
+\frac{3}{160}(\la\ga_{(3)}\la)(\ga_{a}^{(3)}\mu)
\\
\Delta T_{a\alpha}&:=-\widetilde{T}_{a\alpha}-\frac{3i}{20}(\ga_a^{(1)} \nabla_{(1)} \la)-\frac{1}{5}L_{(2)}(\ga_{a}^{(2)}\mu)
-\frac{2}{5}L_{(4)}(\ga_{a}^{(4)}\mu)\\
&+\frac{1}{5}K_{(1)}(\ga_{a}^{(1)}\la)+\frac{3}{20}K_{(3)}(\ga_{a}^{(3)}\la)
+\frac{3}{160}(\mu\ga_{(3)}\mu)(\ga_{a}^{(3)}\la)
~,}}
cf.~(4.9), (4.10) of \cite{Tsimpis:2005vu}.  
Therefore the right-hand side of the Einstein equation (\ref{eeq}) can be modified by a term of the form, 
%
%\eq{
$c_5(\Delta T_{(b}\gamma_{c)}\la)+c_6(\Delta T_{(b}\gamma_{c)}\mu)+
c_7\eta_{bc}(\Delta T\la)+c_8\eta_{bc}(\Delta T\mu)$. 
%~.}
%
Demanding that the Einstein equation thus modified agrees with the Einstein equation coming from (\ref{action2}) leads to 
a highly overdetermined system of equations. As required for consistency, a unique solution exists and is given by,
\eq{
c_5=c_6=-24~;~~~c_7=c_8=-\frac{81}{4} 
~.}

\section{General dilatonic vacua}\label{sec:gen}

The dilatino $\psi_m$ of the superspace formulation is canonically related (through the suspersymmetry transformations) to the metric $g_{mn}$, whereas the dilatino $\Psi_m$ of (\ref{gre}) is canonically related to the rescaled Einstein-frame metric $\hat{g}_{mn}$, cf.~(\ref{rsm}). The action (\ref{action2}) is obtained by setting the 
superspace gravitino to zero which thus corresponds to,
\eq{\label{fv}{}\psi_m{}\equiv0\leftrightarrow {}\Psi_m{}\equiv-\frac34\Gamma_m{}\Lambda{}~,}
as can be seen from (\ref{gfd}).

More generally, setting the  gravitino to zero is a frame-dependent statement. This can be seen directly from the supersymmetry transformation for the vielbein (\ref{vst}) 
which, when evaluated at the lowest order in the $\theta$-expansion, gives $\delta_{\xi}e_m{}^a=-i(\epsilon\ga^a\psi_m)-i(\zeta\ga^a\psi_m)$, up to a Lorentz transformation. 
More generally, it canonically associates the vielbein of the metric $g^{(\beta)}$ with the 
gravitino $\psi^{(\beta)}$, where,
\eq{\label{frames}g^{(\beta)}_{mn}:= e^{2\beta\phi}\hat{g}_{mn}~;~~~
\psi^{(\beta)}_m:=\Psi_m- \beta\!~\Gamma_m{}\Lambda{}~;~~~\beta\in\mathbb{R}~,
}
and we have used, $\delta_{\xi}\phi=\xi\cdot\nabla\phi=(\epsilon\lambda)+(\zeta\mu)$. It follows that setting the gravitino  $\psi^{(\beta)}$ to zero corresponds to,
\eq{\label{fvgen}{}\psi^{(\beta)}_m{}\equiv0\leftrightarrow {}\Psi_m{}\equiv\beta\Gamma_m{}\Lambda{}~,}
which generalizes (\ref{fv}) to an arbitrary frame. In particular,  we distinguish the following cases,
\eq{\label{34}
\beta=\left\{\begin{array}{rl}
-\frac{3}{4},&~~~\text{vanishing~superspace-frame~gravitino} \\
0,&~~~\text{vanishing~Einstein-frame~gravitino}\\
\frac14, &~~~\text{vanishing~string-frame~gravitino}~.
\end{array}\right.
}
The four-fermion part of the 
IIA Lagrangian in \cite{Giani:1984wc} is
given as a sum of 24 terms expressed in terms of $\widehat{\Psi}^{GP}_m$, $\lambda^{GP}$, cf.~appendix \ref{sec:comparison}.  
Substituting (\ref{fvgen}) in \cite{Giani:1984wc} corresponds to setting, 
\eq{\label{fvgpt}{}\psi^{GP}_m{}\equiv \beta\sqrt{2}~\!\Gamma_{11}\Gamma_{m}{}\lambda^{GP}{}~, 
~~~{}\widehat{\Psi}^{GP}_m{}\equiv c~\!\Gamma_{11}\Gamma_{m}\lambda^{GP}{}~,}
where $c:=\sqrt{2}(\beta+1/12)$, with $\beta\in\mathbb{R}$. 
We thus obtain the following expression for the $(\bar{\lambda}\lambda)^2$ term in \cite{Giani:1984wc},
\eq{\spl{\label{qt}(\bar{\la}\Gamma_{mn}&\Gamma_{11}\la)^2(\frac{26\sqrt{2}}{3}c^3-\frac{29}{4}c^4)
+(\bar{\la}\Gamma_{mnpq}\la)^2(\frac{1}{\sqrt{2}}c^3-\frac{21}{8}c^4)\\
&+(\bar{\la}\Gamma_{mnp}\la)^2(\frac{7}{3\sqrt{2}}c^3-5c^4)
+(\bar{\la}\Gamma_{mnp}\Gamma_{11}\la)^2(-\frac{{2}}{3}c^2+\frac{7}{\sqrt{2}}c^3+\sqrt{2}c^3-6c^4)\\
&=(32c^2-276\sqrt{2}c^3+\frac{1773}{2}c^4)(\bar{\la}\la)^2
~,}}
where in the last equality we used the following Fierz identities,
\eq{\spl{\label{fids}
(\bar{\la}\Gamma_{mn}\Gamma_{11}\la)^2&= 6(\bar{\la}\la)^2\\
(\bar{\la}\Gamma_{mnp}\la)^2&= 48(\bar{\la}\la)^2\\
(\bar{\la}\Gamma_{mnp}\Gamma_{11}\la)^2&= -48(\bar{\la}\la)^2\\
(\bar{\la}\Gamma_{mnpq}\la)^2&=-336(\bar{\la}\la)^2
~.}}
Furthermore substituting (\ref{fvgen}) in the massive IIA action of \cite{Romans:1985tz}, completing it with the quartic-fermion term 
(\ref{qt}) and normalizing to our conventions, cf.~appendix \ref{sec:comparison}, we obtain the one-parameter family 
of dilatonic-condensate pseudoactions,  
\boxedeq{\spl{\label{action3}
S&=S_b+
\int\d^{10}x\sqrt{\hat{g}} \Big\{(1-144\beta^2)
(\bar{\Lambda}\Gamma^m\nabla_m\Lambda)
-(36\beta^2-{10}\beta+\frac{21}{20})e^{5\phi/4}m(\bar{\Lambda}\Lambda)\\
&-\frac{1}{2}(29\beta^2-\frac{9}{{2}}\beta+\frac{5}{16})e^{3\phi/4} F_{mn}(\bar{\Lambda}\Gamma^{mn}\Gamma_{11}\Lambda)
-(4\beta^2+\frac{1}{3}\beta)e^{-\phi/2} H_{mnp}(\bar{\Lambda}\Gamma^{mnp}\Gamma_{11}\Lambda)\\
&-\frac{1}{24}(21\beta^2-\frac{1}{{2}}\beta-\frac{3}{16})e^{\phi/4} G_{mnpq}(\bar{\Lambda}\Gamma^{mnpq}\Lambda)
-(8c^2-69\sqrt{2}c^3+\frac{1773}{8}c^4)(\bar{\Lambda}\Lambda)^2
\Big\}
~,}}
where $S_b$ is given in (\ref{ba}), and $c$ was defined below (\ref{fvgpt}). 

Setting $\beta=-3/4$ in (\ref{action3})  
we recover the action (\ref{action2}). 
The dilatonic-condensate pseudoactions $S^E$, $S^{st}$ obtained by setting the Einstein-frame,  string-frame gravitino to zero ($\beta=0,1/4$ respectively) read,
\eq{\spl{\label{action3e}
S^E=S_b+
\int\d^{10}x\sqrt{\hat{g}} \Big\{ 
&(\bar{\Lambda}\Gamma^m\nabla_m\Lambda)
-\frac{21}{20}e^{5\phi/4}m(\bar{\Lambda}\Lambda)+\frac{3}{512}(\bar{\Lambda}\Lambda)^2\\
&-\frac{5}{32}e^{3\phi/4} F_{mn}(\bar{\Lambda}\Gamma^{mn}\Gamma_{11}\Lambda)
+\frac{1}{128}e^{\phi/4} G_{mnpq}(\bar{\Lambda}\Gamma^{mnpq}\Lambda)
\Big\}
~,}}
and
\eq{\spl{\label{action3s}
S^{st}&=S_b+
\int\d^{10}x\sqrt{\hat{g}} \Big\{-8
(\bar{\Lambda}\Gamma^m\nabla_m\Lambda)
-\frac{4}{5}e^{5\phi/4}m(\bar{\Lambda}\Lambda)   -\frac{5}{2}(\bar{\Lambda}\Lambda)^2\\
&-\frac{1}{2} e^{3\phi/4} F_{mn}(\bar{\Lambda}\Gamma^{mn}\Gamma_{11}\Lambda)
-\frac{1}{3}e^{-\phi/2} H_{mnp}(\bar{\Lambda}\Gamma^{mnp}\Gamma_{11}\Lambda)
-\frac{1}{24} e^{\phi/4} G_{mnpq}(\bar{\Lambda}\Gamma^{mnpq}\Lambda)
\Big\}
~.}}
Note that the quartic-dilaton term in $S^E$ can potentially generate a positive cosmological constant, contrary to the quartic-dilaton term in $S^{st}$, which is 
negative.

The dilaton and Einstein equations following from action (\ref{action3}) read,
\eq{\spl{\label{beomf1}
0&=-\hat{\nabla}^2\phi+\frac{3}{8}e^{3\phi/2}F^2-\frac{1}{12}e^{-\phi}H^2+\frac{1}{96}e^{\phi/2}G^2 +\frac{4}{5}m^2e^{5\phi/2}\\
&-\frac{5}{4}(36\beta^2-{10}\beta+\frac{21}{20})e^{5\phi/4}m(\bar{\Lambda}\Lambda)
-\frac{3}{8}(29\beta^2-\frac{9}{{2}}\beta+\frac{5}{16})e^{3\phi/4} F_{mn}(\bar{\Lambda}\Gamma^{mn}\Gamma_{11}\Lambda)\\
&+\frac12 (4\beta^2+\frac{1}{3}\beta)e^{-\phi/2} H_{mnp}(\bar{\Lambda}\Gamma^{mnp}\Gamma_{11}\Lambda)
-\frac{1}{96}(21\beta^2-\frac{1}{{2}}\beta-\frac{3}{16})e^{\phi/4} G_{mnpq}(\bar{\Lambda}\Gamma^{mnpq}\Lambda)
~,
}}
\vfill\break

and,
\eq{\spl{\label{beomf2}
0&=\hat{R}_{mn}+\frac{1}{2}\partial_m\phi\partial_n\phi+\frac{1}{25}m^2e^{5\phi/2}\hat{g}_{mn}
+\frac{1}{4}e^{3\phi/2}\Big(  2F^2_{mn} -\frac{1}{8} \hat{g}_{mn}  F^2 \Big)\\
&+\frac{1}{12}e^{-\phi}\Big(  3H^2_{mn} -\frac{1}{4} \hat{g}_{mn}  H^2 \Big)
+\frac{1}{48}e^{\phi/2}\Big(   4G^2_{mn} -\frac{3}{8} \hat{g}_{mn}  G^2 \Big)\\
&+(1-144\beta^2)\Big(
\frac12(\bar{\Lambda}\Gamma_{(m}\nabla_{n)}\Lambda)
+\frac{1}{16}g_{mn}(\bar{\Lambda}\Gamma^{i}\nabla_{i}\Lambda)
\Big)
\\
&-\frac18 \hat{g}_{mn}\Big( (36\beta^2-{10}\beta+\frac{21}{20})e^{5\phi/4}m(\bar{\Lambda}\Lambda)
+(8c^2-69\sqrt{2}c^3+\frac{1773}{8}c^4)(\bar{\Lambda}\Lambda)^2
\Big)\\
&-\frac{1}{2}(29\beta^2-\frac{9}{{2}}\beta+\frac{5}{16})e^{3\phi/4} 
F_{(m}{}^i(\bar{\Lambda}\Gamma_{n)i}\Gamma_{11}\Lambda)
\\
&-(4\beta^2+\frac{1}{3}\beta)e^{-\phi/2} 
\Big(
\frac32 H_{(m}{}^{ij}(\bar{\Lambda}\Gamma_{n)ij}\Gamma_{11}\Lambda)
-\frac{1}{16} \hat{g}_{mn}H_{(3)}(\bar{\Lambda}\Gamma^{(3)}\Gamma_{11}\Lambda)
\Big)\\
&-\frac{1}{24}(21\beta^2-\frac{1}{{2}}\beta-\frac{3}{16})e^{\phi/4} 
\Big(
2G_{(m}{}^{ijk}(\bar{\Lambda}\Gamma_{n)ijk}\Lambda)
-\frac18 \hat{g}_{mn}
G_{(4)}(\bar{\Lambda}\Gamma^{(4)}\Lambda)
\Big)
~.
}}
The form equations read,
\eq{\spl{\label{beomf3}
0&=\d\Big(\hat{\star}\big[ e^{3\phi/2}F  
-(29\beta^2-\frac{9}{{2}}\beta+\frac{5}{16})e^{3\phi/4} (\bar{\Lambda}\Gamma_{(2)}\Gamma_{11}\Lambda)
\big]\Big)+e^{\phi/2}H\swed \hat{\star} G\\
0&=\d\Big(\hat{\star} 
\big[
e^{-\phi}H
-(24\beta^2+2\beta)e^{-\phi/2} (\bar{\Lambda}\Gamma_{(3)}\Gamma_{11}\Lambda)
\big]
\Big)
+e^{\phi/2}F\swed\hat{\star} G-\frac{1}{2}G\swed G+\frac{4}{5}m e^{3\phi/2}\hat{\star} F\\
0&=\d\Big(
\hat{\star} 
\big[
e^{\phi/2}G
-(21\beta^2-\frac{1}{{2}}\beta-\frac{3}{16})e^{\phi/4} (\bar{\Lambda}\Gamma_{(4)}\Lambda)
\big]
\Big)
-H\swed G
~,
}}
where we have defined: 
$(\bar{\Lambda}\Gamma_{(p)}\Lambda):=\frac{1}{p!} (\bar{\Lambda}\Gamma_{m_1\dots m_p}\Lambda)\d x^{m_p}\wedge\dots\wedge\d x^{m_1}$, 
similarly to our definition for the bosonic forms, cf.~footnote \ref{f2}.

In addition to the equations above, the forms obey the Bianchi identities given in (\ref{bi}).

\section{de Sitter vacua}\label{sec:desitter}

Having obtained the general dilatino-condensate action (\ref{action3}), we can look for de Sitter solutions supported by nonvanishing dilatino condensates. 
We will use for that purpose the dilatino-condensate pseudoaction (\ref{action3e}), obtained by setting the Einstein-frame gravitino to zero ($\beta=0$), although the analysis 
can be easily extended to a general value of the parameter $\beta$.

\subsection{$dS_{10}$}\label{sec:desitter10}

In this section we show that the massless IIA theory admits ten-dimensional de Sitter vacua 
supported by the quartic-dilatino condensate, with constant dilaton and vanishing flux.  
The only potentially nonvanishing condensates in 
the ten-dimensional Lorentz-invariant vacuum are the scalar condensates $(\bar{\Lambda}\Lambda)$ and $(\bar{\Lambda}\Lambda)^2$. 
Note in particular that these  vevs are a priori independent.\footnote{Strictly-speaking 
 these vevs should be denoted by $\langle\bar{\Lambda}\Lambda\rangle$ and $\langle(\bar{\Lambda}\Lambda)^2\rangle$ respectively, where 
 $\langle(\bar{\Lambda}\Lambda)^2\rangle\neq \langle\bar{\Lambda}\Lambda\rangle^2$ in general. 
Omitting the brackets should hopefully not lead to confusion.}

With these assumptions, setting $m, \beta=0$, we see that the Bianchi identities (\ref{bi}), the form equations in (\ref{beomf3})  and the dilaton equation (\ref{beomf1}) are trivially satisfied. 
Moreover the Einstein equation (\ref{beomf2}) reduces to,
\eq{
-\hat{R}_{mn}=\frac{3}{2^{12}}(\bar{\Lambda}\Lambda)^2\hat{g}_{mn}~.
}
For a nonvanishing quartic-dilatino condensate we thus obtain a simple realization of $dS_{10}$ in massless IIA theory.\footnote{\label{f3}Note that 
in our ``superspace'' conventions for the forms, $\hat{R}<0$, $\hat{R}>0$ corresponds to de Sitter, anti-de Sitter space respectively, cf.~also footnote \ref{f2}.} The de Sitter radius is set by the value of the condensate.

\subsection{$dS_{4}\times M_6$ without flux}\label{sec:desitter4sf}

Let us  now consider compactifications, on six-dimensional  K\"{a}hler-Einstein manifolds $M_6$, of massless IIA supergravity to a maximally-symmetric Lorentzian manifold $M_{1,3}$ with 
vanishing flux,  $F$, $H$, $G=0$, and constant dilaton which we also set to zero for simplicity, $\phi=0$. 
More specifically, we assume that the ten-dimensional spacetime is of direct product form $M_{1,3}\times M_6$, 
\eq{\label{dpst}\d s^2=\d s^2(M_{1,3})+ \d s^2(M_{6})
~.}
Moreover,
\eq{\label{ei43}
-{R}_{\mu\nu}= \Omega ~\!g_{\mu\nu}~;~~~
-{R}_{mn}= \omega g_{mn}
 ~,}
where  $g_{\mu\nu}$,   $g_{mn}$ are the components of the metric in the 
external, internal space respectively; we have chosen the parameterization so that positive 
$\Omega$  corresponds to de Sitter space, and similarly for  $\omega$, cf.~footnote \ref{f3}.

The internal manifold being  K\"{a}hler-Einstein, it admits a nowhere-vanishing  spinor, $\eta$, of 
positive chirality, which we take to be commuting.  Moreover  the spinor obeys,
\eq{\nabla_{m}\eta=i\mathcal{P}_m\eta~,}
where $\d \mathcal{P}$ is proportional to $J$, the  K\"{a}hler form of $M_6$. Furthermore $J$ can be expressed as 
an $\eta$ bilinear,
\eq{i\eta^{\dagger}\gamma_{(2)}\eta=J~.}
We decompose the chiral and antichiral components of the dilatino, $\lambda$ and $\mu$ respectively, cf.~(\ref{gfd}), as follows,
\eq{\label{sdec}
\lambda=\chi_+\otimes\eta+c.c.~;~~~ \mu=\chi_-\otimes\eta+c.c.
~,}
where $\chi_{+}$ ($\chi_{-}$) is a  chiral (antichiral) anticommuting Weyl spinor of $M_{1,3}$. The rationale for  this 
decomposition is that, in the 
effective action describing the compactification  on $M_6$,  (\ref{sdec}) should give rise to ``light'' four-dimensional spinors 
$\chi_{\pm}$;\footnote{Although certainly plausible, this is hard to show in general beyond the Calabi-Yau case.} it generalizes to the  K\"{a}hler-Einstein case  the decomposition of \cite{Dine:1985rz}, where $M_6$ is taken to be a Calabi-Yau. 
Similar decompositions were adopted in e.g.~\cite{Gemmer:2013ica}.

It follows from (\ref{sdec}) that, for a  Lorentz-invariant  four-dimensional vacuum, the dilatino bilinear condensates take the form,
\eq{
 (\bar{\Lambda}\Lambda)=\Re (A)~;~~~
 (\bar{\Lambda}\Gamma_{(2)}\Lambda)=\Re(A)~\!J~;~~~
 (\bar{\Lambda}\Gamma_{(4)}\Lambda)=\Im (A)~\! \mathrm{vol}_4+  \Re (A)~\!  \frac12 J^2
~,}
where the complex number $A:= 4(\bar{\chi}_+\chi_-)$ is the four-dimensional quadratic-dilatino condensate,  and 
$\mathrm{vol}_4$ is the volume element of $M_{1,3}$.  
Furthermore, setting $m, \beta=0$, we see that 
the Bianchi identities (\ref{bi}),  the form equations (\ref{beomf3}) and the dilaton equation 
 (\ref{beomf1}) 
are all automatically satisfied. The mixed $(\mu,m)$  components of the Einstein equations (\ref{beomf2}) are automatically satisfied, 
while 
the internal and external components 
of the Einstein equations reduce to,
\eq{\label{esf24}
\Omega=\omega=\frac{3}{2^{12}}(\bar{\Lambda}\Lambda)^2
~,}
where we have used that  vevs of the form $(\bar{\Lambda} \Gamma_{(m}\nabla_{n)}\Lambda)$ vanish. 

For a nonvanishing quartic-dilatino condensate we thus obtain a simple realization of $dS_{4}\times M_6$ in massless IIA theory. 
The curvature of  de Sitter space and the internal manifold are both set by the value of the condensate.

\subsection{$dS_{4}\times M_6$ with RR flux}\label{sec:desitter4}

In this section we consider compactifications   supported solely by a nonvanishing quartic-dilaton vev, i.e. 
such that,
\eq{
(\bar{\Lambda}\Lambda)^2\neq0~; ~~~ (\bar{\Lambda}\Gamma_{(p)}\Lambda)=0
~,}
for $p=0,\dots,10$. 
As we will see, with this assumption\footnote{Lattice models with nonvanishing quartic-fermion condensates and vanishing quadratic-fermion condensates have been studied  in e.g. \cite{Ayyar:2016lxq}.} 
the theory admits four-dimensional de Sitter solutions of the form $dS_4\times M_6$ with nonvanishing RR flux, where 
$M_6$ is a six-dimensional K\"{a}hler-Einstein manifold of positive scalar curvature.

Let us note that solutions with vanishing 
quadratic condensates, supported by nonvanishing quartic-fermion condensates, are necessarily nonsupersymmetric. This 
readily follows from the supersymmetric integrability theorem for (massive) IIA \cite{Tsimpis:2005vu,Lust:2008zd}. Indeed quartic condensates leave the Bianchi identities, the form equations and 
the supersymmetry transformations unchanged, while modifying the Einstein equation. If such 
solutions were supersymmetric they would therefore violate the integrability theorem, leading to contradiction. 
A supersymmetric integrability theorem in the presence of condensates has recently been presented in \cite{Minasian:2017eur} in the context of the 
heterotic string, and it would be interesting to extend it to the type II case.

As in section \ref{sec:desitter4sf}, we assume that the ten-dimensional spacetime is of direct product form, cf.~(\ref{dpst}), (\ref{ei43}). 
Moreover we set the dilaton and the three-form flux to zero, $\phi=0$, $H=0$, and we parameterize the RR fluxes as follows,
\eq{
F=b J~;~~~G=a~\!\mathrm{vol}_4+\frac12 c J^2~;~~~a,b,c\in\mathbb{R}
~,}
where $J$ is the K\"{a}hler form of $M_6$, and $\mathrm{vol}_4$ is the volume element of $M_{1,3}$.  
It is then straightforward to 
see that the Bianich identities (\ref{bi}),  the $F$-form and $G$-form equations in (\ref{beomf3}) are automatically satisfied, while the $H$-form equation reduces to,
\eq{\label{e21}
bc-\frac12 ac+\frac25 mb=0~.}
Moreover the dilaton equation (\ref{beomf1}) reduces to,
\eq{\label{e22}
a^2=9b^2+3c^2+\frac{16}{5}m^2~.}
The two equations above can be used to determine two of the parameters $a$, $b$, $c$, $m$ in terms of the other two. 

The mixed $(\mu,m)$  components of the Einstein equations (\ref{beomf2}) are automatically satisfied, 
while 
the internal components 
of the Einstein equations reduce to,
\eq{\label{e24}
\omega=\frac{16}{25}m^2+2b^2+c^2+\frac{3}{2^{12}}(\bar{\Lambda}\Lambda)^2
~,}
where we have taken (\ref{e22}) into account. 
This equation simply solves for $\omega$;  
it implies that the internal space $M_6$ is necessarily of positive scalar curvature.

Lastly the $(\mu,\nu)$ components of the Einstein equations reduce to,
\eq{\label{e23}
\Omega=-3b^2-\frac32c^2-\frac{24}{25}m^2+\frac{3}{2^{12}}(\bar{\Lambda}\Lambda)^2
~,}
where again we have used (\ref{e22}). 
It follows that for $(\bar{\Lambda}\Lambda)^2$ sufficiently large, $\Omega$ is positive and 
the theory admits $dS_4\times M_6$ solutions. We also note that, for vanishing condensate,  
$\Omega$ is necessarily negative. In this case we recover 
the $AdS_4\times M_6$ solutions described in section 3.2 of \cite{Lust:2009zb}.

\section{Conclusions}\label{sec:conclusions}

We have used the superspace formulation \cite{Tsimpis:2005vu} of Romans supergravity \cite{Romans:1985tz} to obtain the dilatino terms of the theory, and we have found agreement with the quartic-fermion term of \cite{Giani:1984wc}.  
As we have seen, setting the Einstein-frame gravitino to zero results in a positive quartic-dilatino 
term, which could therefore  generate a positive cosmological constant via fermionic condensation.

As a byproduct we have obtained the superform formulation of Romans supergravity: the hatted superforms of eq.~(\ref{hs}) obey the 
 super-Bianchi identites (\ref{214}). The latter can be used as an alternative starting point for defining the full theory in superspace.

We have shown that the theory admits formal de Sitter space solutions, obtained by assuming nonvanishing dilatino condensates.   
This is in contrast to gaugino-condensate scenarios in heterotic string which do not seem to allow for a de Sitter vacuum \cite{Quigley:2015jia}. 
The results of the present paper open the way for a more general and systematic study of (massive) IIA solutions supported by dilatino condensates, with or without supersymmetry.

We emphasize that we { do not} claim to have solved the problem of de Sitter space in string theory: we have offered neither a concrete mechanism for the generation of the dilatino condensate (although we have brane  instantons in mind), nor any controlled setting in which these quantum effects might take place.~The main message of the present paper is that the quartic-dilatino term in 
(massive) IIA turns out to be positive, and that this could potentially be important for cosmological applications.

It is well known that de Sitter and, more generally, cosmological spacetimes are not straightforward to embed in string theory. 
In that respect fermionic condensates offer an interesting possibility for generating 
a positive cosmological constant. Although 
elucidating the quantum origin of the putative dilatino condensate is beyond the scope of the present paper, it 
is clearly an important point that needs to be addressed.

\section*{Acknowledgment} 

We are grateful to Stefan Theisen for related collaboration during July-September 2013.

\appendix

\section{The supersymmetry transformations}\label{sec:sse}

Although we do not directly make use of this in the present paper, it is instructive to work out the explicit form of the supersymmetry 
transformations. A superdiffeomorphism generated by the supervector field $\xi^A$ acts on the vielbein as follows,
\eq{\label{vst}
\delta_{\xi} E_M{}^A=\nabla_M\xi^A+\xi^BT_{BM}{}^A
~,}
up to a $\xi$-dependent Lorentz transformation.  
The supersymmetry transformation of the 
gravitini, $\psi_m^{\alpha}:=\left.E_m^{\alpha}\right|$, $\psi_{m\alpha}:=\left.E_{m\alpha}\right|$, with parameters $(\epsilon^{\alpha}, \zeta_{\alpha})$, 
is obtained from the above 
by setting $\epsilon^{\alpha}:=\xi^{\alpha}|$,  $\zeta_{\alpha}:=\xi_{\alpha}|$, 
where the vertical bar denotes the lowest-order term in the theta-expansion. We thus obtain,
\eq{\spl{
\delta\psi_m^{\alpha}&=
\nabla_m\epsilon^{\alpha}+e_m{}^c(\epsilon^{\beta}T_{\beta c}{}^{\alpha} +\zeta_{\beta}T^{\beta}{}_{c}{}^{\alpha})|\\
\delta\psi_{m\alpha}&=
\nabla_m\zeta_{\alpha}+e_m{}^c(\epsilon^{\beta}T_{\beta c\alpha} +\zeta_{\beta}T^{\beta}{}_{c\alpha})|
~,}}
up to gravitino-dependent, cubic fermion terms which we do not need to consider here. 
Correspondingly the supersymmetry transformation of the dilatini reads,
\eq{\spl{\label{ssy}
\delta\mu^{\alpha}&=(\epsilon^{\beta}\nabla_{\beta}\mu^{\alpha}+\zeta_{\beta}\nabla^{\beta}\mu^{\alpha})| \\
&=L\epsilon+K_m{\gamma}^m\zeta
-L_{mn}{\gamma}^{mn}\epsilon
+K_{mnp}{\gamma}^{mnp}\zeta
+L_{mnpq}{\gamma}^{mnpq}\epsilon
\\
\delta\lambda_{\alpha}&=(\epsilon^{\beta}\nabla_{\beta}\lambda_{\alpha}+\zeta_{\beta}\nabla^{\beta}\lambda_{\alpha})|\\
&=-L\zeta+K_m{\gamma}^m\epsilon
-L_{mn}{\gamma}^{mn}\zeta
-K_{mnp}{\gamma}^{mnp}\epsilon
-L_{mnpq}{\gamma}^{mnpq}\zeta
~,}}
where we have taken (4.5),(4.6) of \cite{Tsimpis:2005vu} into account. Together with (\ref{sc}),(\ref{dic}) above we obtain, suppressing spinor indices,
\eq{\spl{\label{dils1}
e^{-3\phi/4}\delta\mu&= 
\frac{i}{2}\partial_m\phi\hat{\gamma}^m\zeta+\frac12 m e^{5\phi/4}\epsilon\\
&+\frac{3}{16}e^{3\phi/4}F_{mn}\hat{\gamma}^{mn}\epsilon
-\frac{i}{24}e^{-\phi/2}H_{mnp}\hat{\gamma}^{mnp}\zeta
+\frac{1}{192}e^{\phi/4}G_{mnpq}\hat{\gamma}^{mnpq}\epsilon
\\
e^{-3\phi/4}\delta\lambda&= 
\frac{i}{2}\partial_m\phi\hat{\gamma}^m\epsilon-\frac12 m e^{5\phi/4}\zeta\\
&+\frac{3}{16}e^{3\phi/4}F_{mn}\hat{\gamma}^{mn}\zeta
+\frac{i}{24}e^{-\phi/2}H_{mnp}\hat{\gamma}^{mnp}\epsilon
-\frac{1}{192}e^{\phi/4}G_{mnpq}\hat{\gamma}^{mnpq}\zeta
~,}}
up to cubic fermion terms; the curved gamma matrices $\hat{\gamma}$ are defined with respect to the 
rescaled metric (\ref{rsm}).  
Similarly for the gravitino transformations we obtain,
\eq{\spl{\label{gravs1p}
\delta\psi_{m\alpha}&={\nabla}_m\zeta-S{\gamma}_m\epsilon
+F^1_{ef}\gamma_m{}^{ef}\epsilon
-F^2_{me}\gamma^{e}\epsilon\\
&-H^{\prime1}_{fgh}\gamma_m{}^{fgh}\zeta+H^{\prime2}_{mgh}\gamma^{gh}\zeta
-G^1_{efgh}\gamma_m{}^{efgh}\epsilon+G^2_{mefg}\gamma^{efg}\epsilon
\\
\delta\psi_m^{\alpha}&={\nabla}_m\epsilon+S{\gamma}_m\zeta
+F^1_{ef}\gamma_m{}^{ef}\zeta
-F^2_{me}\gamma^{e}\zeta\\
&-H^1_{fgh}\gamma_m{}^{fgh}\epsilon+H^2_{mgh}\gamma^{gh}\epsilon
+G^1_{efgh}\gamma_m{}^{efgh}\zeta-G^2_{mefg}\gamma^{efg}\zeta
~,}}
where we used (4.3) of \cite{Tsimpis:2005vu}. Furthermore using (4.6) of \cite{Tsimpis:2005vu} and (\ref{sc}),(\ref{dic}) above we obtain, 
\eq{\spl{\label{gravs1}
\delta\psi_{m\alpha}&=\hat{\nabla}_m\zeta+\frac{2i}{5}me^{5\phi/4}\hat{\gamma}_m\epsilon
+\frac38\partial_e\phi\hat{\gamma}^e{}_{m}\zeta+\frac{i}{8}e^{3\phi/4}F_{ef}\gamma_m{}^{ef}\epsilon
+\frac{i}{2}e^{3\phi/4}F_{me}\gamma^{e}\epsilon\\
&+\frac{1}{24}e^{-\phi/2}H_{fgh}\gamma_m{}^{fgh}\zeta
+\frac{i}{24}e^{\phi/4}G_{mefg}\gamma^{efg}\epsilon
\\
\delta\psi_m^{\alpha}&=\hat{\nabla}_m\epsilon-\frac{2i}{5}me^{5\phi/4}\hat{\gamma}_m\zeta
+\frac38\partial_e\phi\hat{\gamma}^e{}_{m}\epsilon+\frac{i}{8}e^{3\phi/4}F_{ef}\gamma_m{}^{ef}\zeta
+\frac{i}{2}e^{3\phi/4}F_{me}\gamma^{e}\zeta\\
&-\frac{1}{24}e^{-\phi/2}H_{fgh}\gamma_m{}^{fgh}\epsilon
-\frac{i}{24}e^{\phi/4}G_{mefg}\gamma^{efg}\zeta
~,}}
up to cubic fermion terms; $\hat{\nabla}$ is the covariant derivative associated to the spin connection of the rescaled metric (\ref{rsm}) so that,
\eq{
e^{3\phi/2}\omega_{nkm}=\hat{\omega}_{nkm}+\frac34\hat{g}_{nk}\partial_m\phi-\frac34\hat{g}_{nm}\partial_k\phi
~;~~~\nabla_m\chi=\hat{\nabla}_m\chi+\frac38\partial_n\phi(\gamma^n{}_m\chi)
~,}
where $\hat{\omega}$, $\omega$ are the spin connections of $\hat{g}$, $g$ respectively, and $\chi$ is a fermion of either chirality.

To make contact with the supersymmetry transformations as given in e.g.~\cite{Lust:2004ig} we use the following ten-dimensional Dirac-matrix notation:
\eq{\label{gf}
\Gamma_m=\left(\begin{array}{cc}0&-i(\hat{\gamma}_m)_{\alpha\beta}\\
i(\hat{\gamma}_{m})^{\alpha\beta}&0
\end{array}\right)~;~~~\Gamma_{11}=\left(\begin{array}{cc}\delta_{\beta}^{\alpha} &0\\
0&-  \delta^{\beta}_{\alpha}
\end{array}\right)
~;~~~C^{-1}=\left(\begin{array}{cc}0&\delta_{\beta}^{\alpha} \\
-\delta^{\beta}_{\alpha} &0
\end{array}\right)
~,
}
and define the Dirac-Majorana spinors, 
\eq{\label{gfd}
\Psi_m=e^{3\phi/8}\left(\begin{array}{c}\psi_{m\alpha}\\
\psi_m^{\alpha}
\end{array}\right)-\frac34 \Gamma_{m}\Lambda~;~~~
\Lambda=e^{-3\phi/8}\Gamma_{11}\left(\begin{array}{c}\lambda_{\alpha}\\
\mu^{\alpha}
\end{array}\right)~;~~~
\Theta=e^{3\phi/8}\left(\begin{array}{c}   \zeta_{\alpha}   \\
\epsilon^{\alpha}
\end{array}\right)
~,}
which obey the reality conditions $\overline{\Psi}_m={\Psi}_m^{\mathrm{Tr}}C^{-1}$, etc.  
In terms of these, the supersymmetry transformations (\ref{dils1}), (\ref{gravs1}) take the form,
\eq{\label{dile}
\delta\Lambda=\Big\{
-\frac{1}{2}\Gamma^m\hat{\nabla}_m\phi-\frac{me^{5\phi/4}}{2}
+\frac{3e^{3\phi/4}}{16}F_{mn}\Gamma^{mn}\Gamma_{11}
+\frac{e^{-\phi/2}}{24}H_{mnp}\Gamma^{mnp}\Gamma_{11}
-\frac{e^{\phi/4}}{192}G_{mnpq}\Gamma^{mnpq}
\Big\}\Theta~,
}
and
\eq{\spl{\label{gre}
\delta{\Psi}_m&=\Big\{\hat{\nabla}_m
-\frac{me^{5\phi/4}}{40}\Gamma_m
-\frac{e^{3\phi/4}}{64}
F_{np}(\Gamma_m{}^{np}-14\delta_m{}^n\Gamma^p)\Gamma_{11}\\
&+\frac{e^{-\phi/2}}{96}H_{npq}(\Gamma_m{}^{npq}-9\delta_m{}^n\Gamma^{pq})\Gamma_{11}
+\frac{e^{\phi/4}}{256}G_{npqr}(\Gamma_m{}^{npqr}
-\frac{20}{3}\delta_m{}^n\Gamma^{pqr})\Big\}\Theta
~,
}}
respectively, up to cubic fermion terms. These are precisely the supersymmetry transformations expressed  in the conventions of \cite{Lust:2004ig}.

\section{A note on conventions}\label{sec:comparison}

In this section we compare our conventions to those of \cite{Romans:1985tz,Giani:1984wc} . 
The translation between the conventions of the present paper and those of \cite{Lust:2004ig} was explained previously.

The fermionic fields in \cite{Romans:1985tz} are 
related to those in the present paper via,
\eq{\psi_m^{R}=\Psi_m~;~~~\lambda^{R}=\frac{1}{\sqrt{2}}\Lambda
~,}
where the $R$ superscript denotes the fields in that reference. Moreover the bosonic fields of \cite{Romans:1985tz} are 
related to those in the present paper via,
\eq{\spl{\label{dicc}
m^RB_{(2)}&=\frac{1}{2}F_{(2)}~;~~~
G^{R}_{(3)}=\frac{1}{2}H_{(3)}~;~~~
F^R_{(4)}=\frac{1}{2}G_{(4)}\\
m^R&=\frac{4}{5}m
~;~~~\phi^R=-\frac12\phi
~;~~~
{R}^R=-\hat{R}
~.}}
With these field redefinitions it can be seen that at the fermionic vacuum (\ref{fv}) 
the action of \cite{Romans:1985tz} precisely reduces to that given in (\ref{action2}), (\ref{ba}) of the present paper, up to the quartic-fermion term 
which was not computed in \cite{Romans:1985tz}.

On the other hand the quartic-fermion terms are  identical in the massive and massless IIA theories. 
In order to compare with the quartic-fermion terms of massless IIA as given in \cite{Giani:1984wc} we note that, upon setting $k=1$ therein, 
the fermionic $\psi_m^{GP}$, $\lambda^{GP}$ of that reference are related to 
the ones in the present paper via,
\eq{\Psi_m=\frac{1}{\sqrt{2}}\psi^{GP}_m~;~~~\Lambda=-\Gamma_{11}\lambda^{GP}
~.}
%
%and similarly for the bosonic forms, $F^R=\frac{1}{\sqrt{2}}F^{GP}$.  
Thus the fermionic vacuum (\ref{fv}) corresponds to setting, 
\eq{\label{fvgp}{}\psi^{GP}_m{}\equiv-\frac{3}{2\sqrt{2}}\Gamma_{11}\Gamma_{m}{}\lambda^{GP}{}~, 
~~~{}\widehat{\Psi}^{GP}_m{}\equiv -\frac{2\sqrt{2}}{3}\Gamma_{11}\Gamma_{m}{}\lambda^{GP}{}~,}
where $\widehat{\Psi}^{GP}_m:=\psi^{GP}_m+({\sqrt{2}}/{12})\Gamma_{11}\Gamma_{m}\lambda^{GP}$.

%%%%%%%%%%%%%%%%%%%%%%%%%%%%%%%%%%%%%%%%%%%%
%
% Bibliography
%
%%%%%%%%%%%%%%%%%%%%%%%%%%%%%%%%%%%%%%%%%%%%
%\cleardoublepage
%\pagestyle{plain}
%\def\href#1#2{#2}
%\addcontentsline{toc}{chapter}{\sffamily\bfseries Bibliography}

\end{document}